\documentclass[conference,a4paper]{IEEEtran}

\usepackage{hhline}
\usepackage{textgreek}
\usepackage{textcomp}
\usepackage{gensymb}

\usepackage{xcolor}

\ifCLASSINFOpdf
  \usepackage[pdftex]{graphicx}
\else
  \usepackage[dvips]{graphicx}
\fi
\usepackage[cmex10]{amsmath}
\DeclareRobustCommand*{\IEEEauthorrefmark}[1]{\raisebox{0pt}[0pt][0pt]{\textsuperscript{\footnotesize #1}}}

\begin{document}
%
\title{Directional Wideband Channel Measurements at 28\,GHz in an Industrial Environment}

\author{\IEEEauthorblockN{
Mathis Schmieder\IEEEauthorrefmark{1},  
Fabian Undi\IEEEauthorrefmark{1},  
Michael Peter\IEEEauthorrefmark{1},  
Ephraim Koenig\IEEEauthorrefmark{1},  
Wilhelm Keusgen\IEEEauthorrefmark{1}  
}                                     
\IEEEauthorblockA{\IEEEauthorrefmark{1}
Fraunhofer Heinrich Hertz Institute, Berlin, Germany, mathis.schmieder@hhi.fraunhofer.de}
}

\maketitle

\begin{abstract}
With the expected adoption of millimeter wave technologies for industrial communication, it is fundamentally important to properly understand the radio channel characteristics of such environments. This paper presents the setup, scenario and results of a measurement campaign at 28 GHz in a machine hall. The radio channel was measured with a bandwidth of 2 GHz and both large scale parameters and directional information were extracted. Evaluation of the power delay profiles shows that the channel contains dense multipath components. A path loss model is parameterized and blockage losses, RMS delay and angle spreads are evaluated. Comparison with the 3GPP TR 38.901 channel model shows that none of the currently defined scenarios is a fit for industrial settings. This emphasizes the need for a newly defined scenario with an industry specific parameter set. 
\end{abstract}


%
\IEEEpeerreviewmaketitle

\section{Introduction}
We are in the midst of the fourth industrial revolution, which brings both huge opportunities and challenges \cite{draht2014industrie4}. It is supposed to open up new application and value-added areas in the production environment. The backbone for this development is high-performance communication for data exchange between machines, tools and work pieces, providing high reliability, low latency and high throughput. In this context, wireless technologies are playing an increasingly important role, as they enable greater flexibility in production driven by individualization ("batch size one") compared to wired transmission and allow the connection of mobile robotics. Fifth generation (5G) mobile network technologies promise to meet these requirements \cite{Wollschlaeger2017}. Companies are planning to set up and operate their own mobile networks at their production sites for this purpose. Initially, such networks will be implemented at frequencies below 6 GHz, but it is foreseeable that the logical expansion of 5G into the millimeter wave range will also advance into the industrial setting.

However, before 5G technologies can be used profitably on a large scale in industrial environments, apart from further technical development, fundamental investigations are necessary to determine the peculiarities of radio propagation in factories to enable proper system parameterization, deployment and performance prediction. For this purpose, the radio channel in such facilities has to be measured, characterized and modeled.

In mobile communications, geometry-based stochastic channel models (GSCMs) have become widely adopted. They allow simulations on link and system level and, among other things, to consider various antenna configurations. The quasi-stochastic empirical models are usually parameterized on the basis of channel measurement data obtained in typical propagation environments. In this context, the GSCMs of the 3rd Generation Partnership Project (3GPP) can be considered as the reference models with the greatest influence. In the past decade, they have been initially established for frequencies around 2 GHz, whereas the latest model targets the full frequency range from 0.5 to 100 GHz \cite{3GPP-TR-38901}. The models and associated channel measurement campaigns, however, focus on the scenarios Rural Macro (RMa), Urban Macro (UMa), Urban Micro (UMi), and indoor office. They do not cover factories. 

For industrial environments, only few results are available in the literature, and the knowledge about the channel is very limited. The early publications \cite{Rappaport89, Rappaport89a} from 1989 present investigations on the large-scale and small-scale characteristics of the channel at 1300 MHz in food processing, engine manufacturing, and aluminium metalwork facilities. The paper \cite{Rappaport89} addresses narrowband path loss and fading behaviour, whereas \cite{Rappaport89a} focuses on the temporal dispersion of the multipath channel. About ten years later, Kiesbu et al. \cite{Kiesbu2000} conducted measurements at 2450 MHz in a chemical pulp factory, a cable factory, and a nuclear power plant to evaluate large-scale and small-scale fading. Tanghe et al. \cite{Tanghe2008} reported narrow-band measurements in two wood processing and two metal processing factories at three frequencies, namely 900 MHz, 2400 MHz, and 5200 MHz. The latest results are available from the National Institute of Standards and Technology (NIST): \cite{Quimby2017} presents analyses of the path gain at 2.245 and 5.4 GHz for three industrial facilities, namely a medium-sized steam plant, an automotive assembly plant and small machine shop. Published results on radio channels in factories for millimeter wave frequencies are so far only based on ray tracing simulations \cite{Solomitckii2018}. In summary, the investigations indicate that the radio channel in factories may behave markedly differently than channels in office buildings due to the open building layout and the presence of machinery and highly reflective materials. The findings stress the need for further measurements to extract accurate model parameters, since they cannot be extrapolated from other environments.

This paper presents first results of wideband channel measurements in a machine hall at 28 GHz yielding directional information on both sides of the link.

\section{Channel Sounder Setup}
\begin{figure}[htb]
    \centering
    \includegraphics[width=0.45\textwidth]{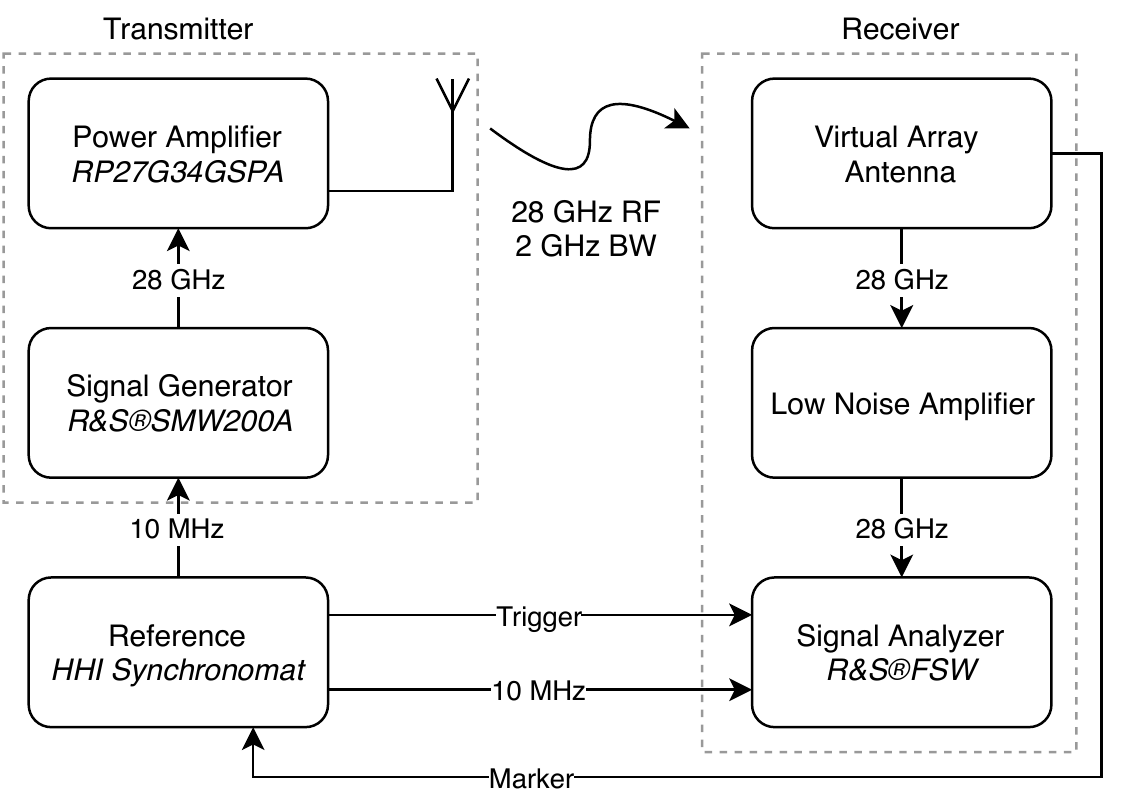}
    \vspace{-0.5em}
    \caption{Channel sounder setup used for measurements}
    \label{fig:setup}
\end{figure}

For the measurements described in this paper, an advanced instrument-based highly flexible time-domain channel sounder was used. The setup is illustrated as a simplified block diagram in Fig. \ref{fig:setup}. Both transmitter (Tx) and receiver (Rx) shared a common reference provided by a high precision rubidium clock  (Synchronomat by \emph{Fraunhofer HHI}), which also enabled coherent triggering at the Rx side. %
\paragraph{Transmitter} %
At the Tx side, a signal generator (R\&S\textsuperscript{\textregistered}SMW200A) was used to generate the channel sounding signal at a frequency of 28\,GHz ($\lambda$ = 10.71\,mm) with a bandwidth of 2\,GHz, resulting in a temporal resolution of $0.$5\,ns. A periodic 96,000 element Frank-Zadoff-Chu (FZC) \cite{chu1972polyphase,frank1973comments} sequence with a duration of 48\,\textmu s, yielding $50.6$\,dB processing gain, was used as sounding signal to estimate the channel transfer function and impulse response. The signal was generated with a power of 19\,dBm, fed through a power amplifier (PA, RF-LAMBDA RP27G34GSPA) and transmitted using a vertically polarized omni-directional antenna. Considering a measured gain of 32\,dB by the PA and cable losses of 17\,dB, the total transmit power at the antenna was 34\,dBm. While the transmit antenna has not yet been fully characterized, simulations imply a $\lambda/4$-monopole performance.
\paragraph{Receiver} %
At the Rx side, a virtual circular array antenna (VCA) \cite{nguyen2016instantaneous} was used. The received signal was fed through a low noise amplifier (LNA, Analog HMC1040) with a gain of 20\,dB and captured using a signal analyzer (R\&S\textsuperscript{\textregistered}FSW). Baseband samples with a resolution of 16\,bit were streamed to a connected PC and stored there. Prior to the measurement campaign, the system was calibrated on site via a back-to-back measurement.

The VCA was used in combination with the real-valued beamspace root-MUSIC algorithm (RB-MUSIC) \cite{lihua2003novel} to extract spatial information from the received signal. The array has a diameter of 97.8\,mm and the rotation speed was set to 1250 rotations per minute (RPM). One rotation and therefore one complete measurement took 48\,ms. By continuously sampling the channel over a full rotation and by choosing a sequence duration of 48\,\textmu s, 1000 virtual antenna elements were equally distributed around the circular aperture. The resulting distance between adjacent virtual array elements was 0.307\,mm (0.0287\,$\lambda$). A zero-position marker is sent from the VCA to the Synchronomat which then triggered the Rx. The channel sounder setup has an instantaneous dynamic range of 65\,dB limited only by thermal noise. The important channel sounder parameters are summarized in Table \ref{table_sounderparameters}.

\begin{table}[ht]
\renewcommand{\arraystretch}{1.0}
\caption{Channel sounder parameters}
\vspace{-1em}
\label{table_sounderparameters}
\centering
\begin{tabular}{|c|c|}
\hline
Type & Value\\
\hhline{|=|=|}
Carrier frequency & 28 GHz\\
\hline
Transmit power & 34 dBm \\
\hline
Sounding bandwidth & 2000 MHz \\
\hline
Sampling rate at Tx & 2400 MHz \\
\hline
Sampling rate at Rx & 2500 MHz \\
\hline
Sequence duration & 48 \textmu s \\
\hline
Temporal snapshot separation & 48 \textmu s \\
\hline
Diameter of virtual array & 97.8 mm \\
\hline
\# of virtual array elements & 1000 \\
\hline
Distance between elements & 0.307 mm (0.0287 $\lambda$) \\
\hline
\end{tabular}
\end{table}
%

\section{Measurement Scenario and Procedure}
\paragraph{Measurement Scenario}%

\begin{figure*}[hbt]
\centering
\includegraphics[width=\textwidth]{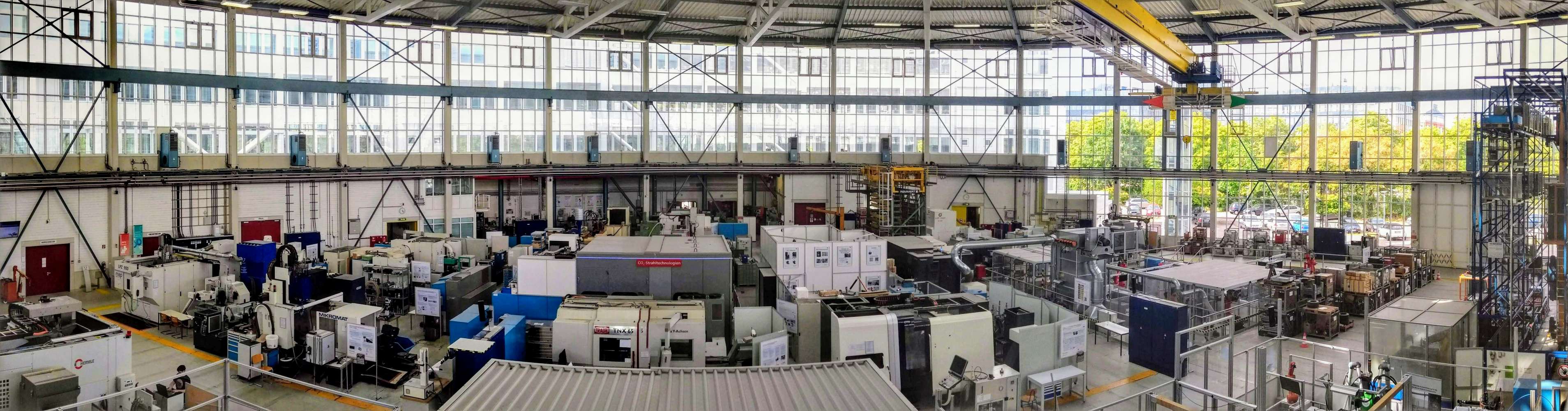}
    \caption{Overview of the measurement scenario from BS position}
    \vspace{-1em}
    \label{fig:hall_picture}
\end{figure*}

\begin{figure}[htb]
    \centering
    \includegraphics[width=0.45\textwidth]{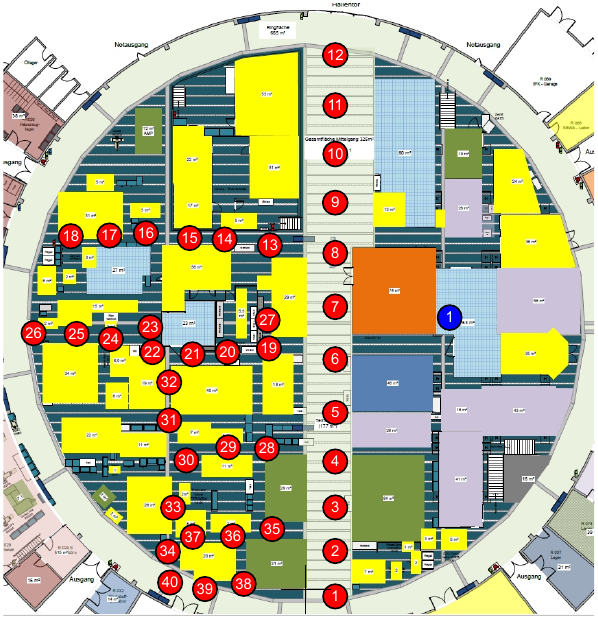}
    \vspace{-0.5em}
    \caption{Floor plan of the machine hall}
    \vspace{-1em}
    \label{fig:scenario}
\end{figure}

The measurements were conducted inside a circle-shaped machine hall with a diameter of 63 m. Surrounding the hall are glass walls interrupted by 40 cm wide concrete pillars every 7.35\,m supporting the dome-shaped metallic roof at a height of 16\,m. In the middle of the hall, the height of the roof rises to 18.5 m. On the ground level, the southern half of the surrounding wall is made of stone, incorporating large metallic doors. Another large metallic door is on the north side of the hall. On the eastern side of the hall, a platform towers in a height of 5\,m above the floor, carrying conference rooms enclosed by glass barriers. The whole floor is packed with industrial machines of various types, predominantly consisting of metal. The hall also features a crane at a height of 12.7\,m. In Fig. \ref{fig:scenario}, the floor plan of the machine hall is illustrated. Fig. \ref{fig:hall_picture} shows the hall as seen from the balcony.

Two types of antenna positions were used in the measurements: A base station (BS) position was emulated by placing the antenna on the balcony at a height of 7\,m, towering above the industrial machines. This position is marked with a blue dot in Fig. \ref{fig:scenario}. The second position type was chosen to emulate user equipment (UE) devices on the floor at heights of 1.7 and 4\,m. Those are marked with red dots in Fig. \ref{fig:scenario}. The minimum distance between Tx and Rx was 11.5\,m, the maximum was 42\,m.

\paragraph{Measurement Procedure}
The measurements have been conducted in two ways: During the first run, the transmitter was placed at the BS position on the balcony while the receiver was positioned at the various UE positions on the floor. In the second run, the transmitter and receiver positions were interchanged in order to gather directional channel information on both sides. The receiver was then placed at the BS position with the transmitter at the various UE positions.
For each UE position, the measurements were performed at two different heights above ground level. In order to ensure a clean line-of-sight (LOS) path between Tx and Rx, the antenna was positioned at 4 meters above ground level, well above the industrial machines. When the antenna height was set to 1.7 meters above ground level, the LOS path was blocked at most positions by machines, therefore creating a non-line-of-sight (NLOS) scenario.

With this procedure, a total of 162 measurements were conducted in both LOS and NLOS conditions and with the receiving VCA at both the BS and UE positions.

\section{Measurement Results and Evaluation}
\begin{figure}[ht]
    \centering
    \includegraphics[width=0.45\textwidth]{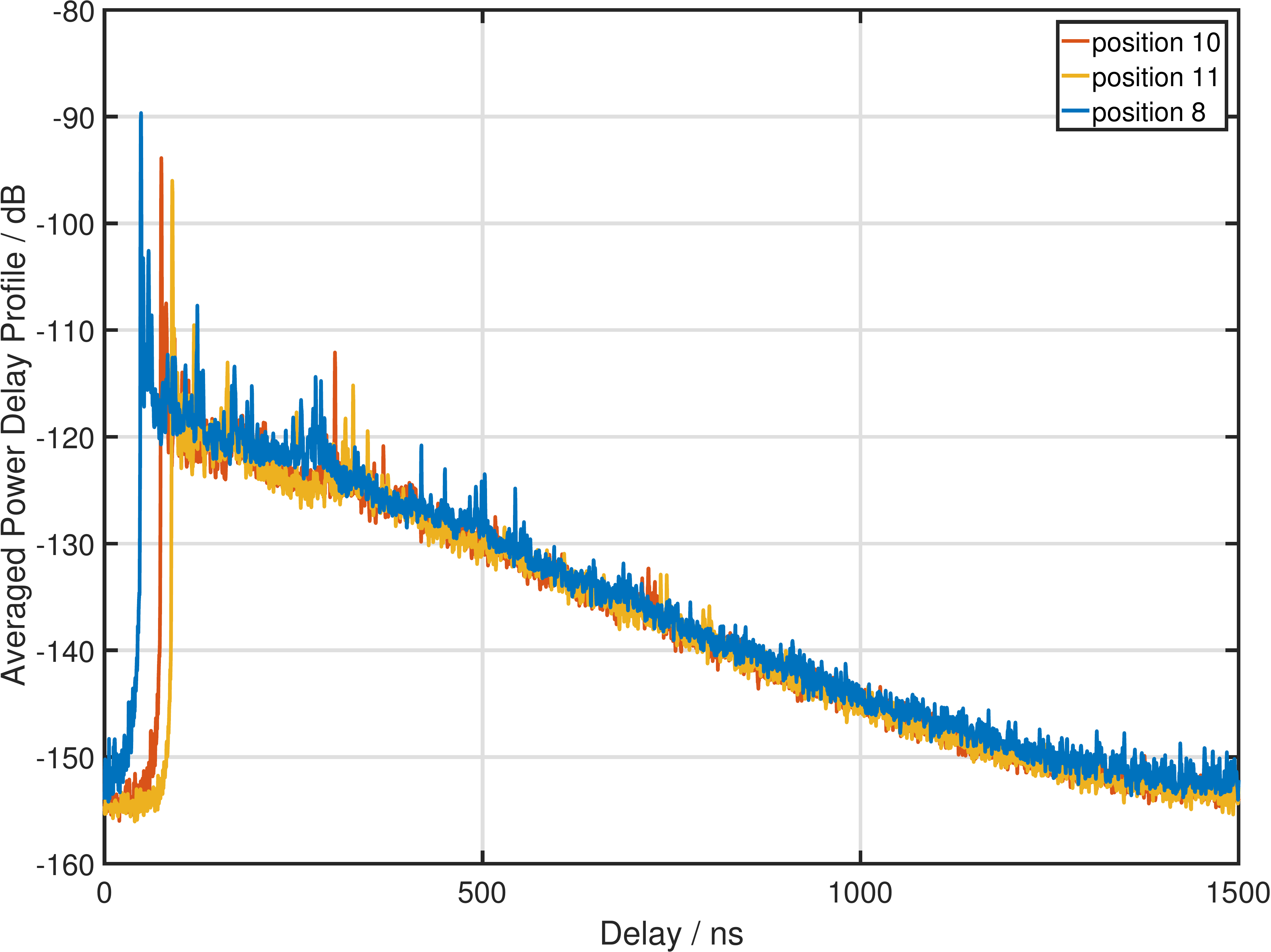}
    \vspace{-0.5em}
    \caption{APDPs for LOS scenarios at three different positions}
    \vspace{-1em}
    \label{fig:apdp1}
\end{figure}

For each of the 162 measurements, the channel sounder produced one channel impulse response (CIR) per virtual antenna array element, resulting in 1000 CIR snapshots per measurement.
The 1000 snapshots were averaged into an average power delay profile (APDP) in order to estimate large scale parameters. Exemplary APDPs for three different measurement points for LOS are illustrated in Fig. \ref{fig:apdp1}. The results show a noise floor of $-155$\,dB, a main LOS component with a power of $-90$ to $-96$\,dB and strong scatterers at delays of about 100 and 300\,ns. Analysis of the APDPs shows that the channel is not sparse and in fact contains dense multipath components (DMCs) \cite{putanen_dmc} up to a delay of more than 1000\,ns until they disappear in thermal noise.

The APDPs were then analyzed for multipath components (MPCs) with respect to an evaluation threshold of 40\,dB relative to the strongest component. Based on the identified MPCs, the path \& blockage loss and delay spread were evaluated. The evaluation was done separately for LOS and NLOS. Additionally, for each of the 162 measurements, the RB-MUSIC algorithm was applied to the 1000 CIR snapshots in order to extract direction-of-arrival (DoA) information. This was then used to estimate the angular spread and power angular spectrum (PAS).

In the following, the channel measurement results are evaluated for path loss, blockage loss and RMS delay and angular spreads. As of yet, no industrial indoor channel model has been agreed on by 3GPP. The scenario with the highest similarity is the \textit{Indoor-Office} scenario described in 3GPP TR 38.901 \cite{3GPP-TR-38901}. The results in this paper are compared to the 3GPP \textit{Indoor-Office} parameters and the findings of Solomitckii et al. \cite{Solomitckii2018} in the 28\,GHz \textit{Heavy Industry} (H28) scenario.

\subsection{Path Loss}
\begin{figure}[ht]
    \centering
    \includegraphics[width=0.45\textwidth]{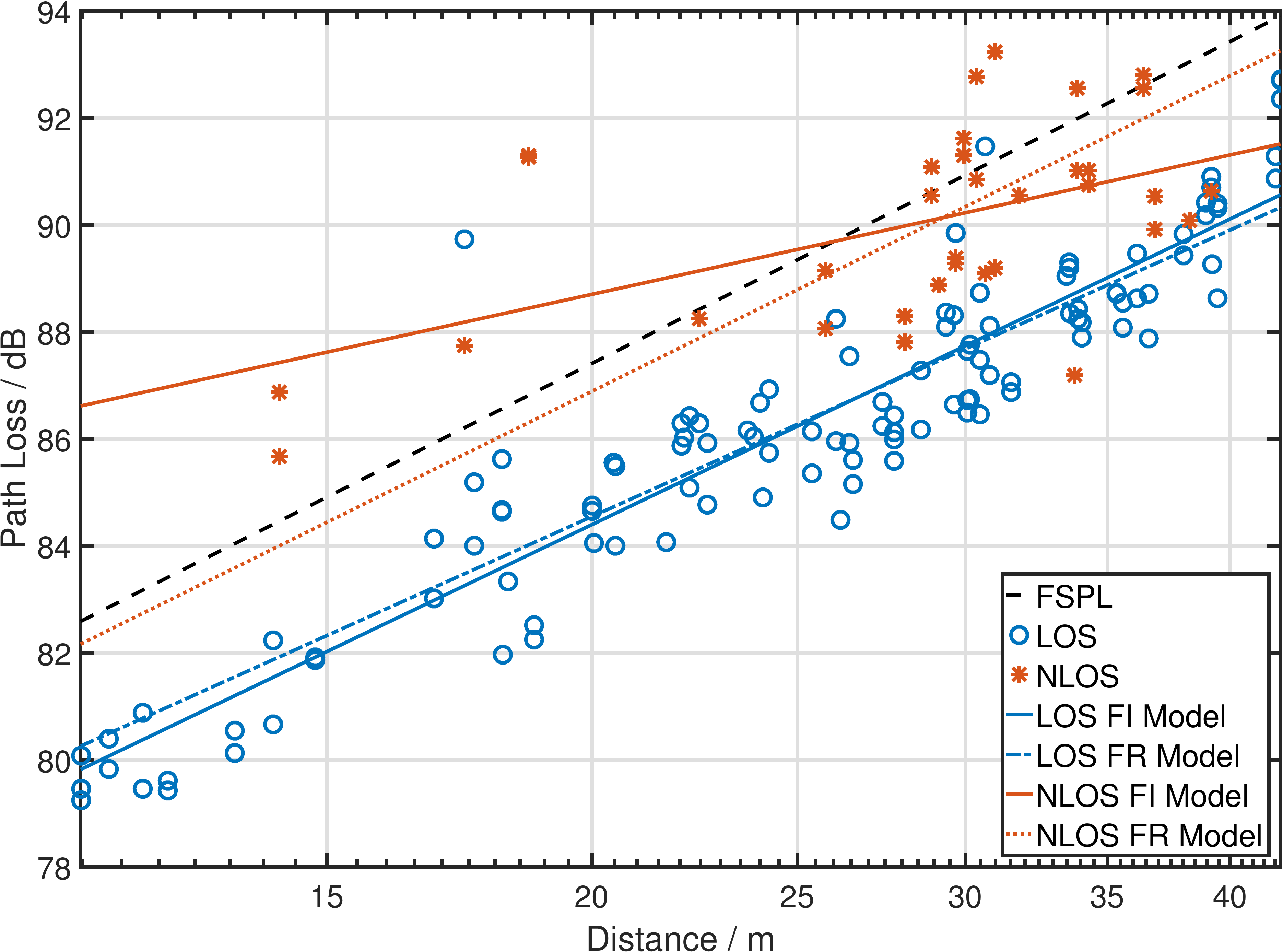}
    \vspace{-0.5em}
    \caption{Path loss in LOS and NLOS condition}
    \vspace{-1em}
    \label{fig:pathloss}
\end{figure}

The results of the path loss evaluation are shown together with the free-space path loss (FSPL) in Fig. \ref{fig:pathloss}. Results from measurements that were classified as LOS are coloured blue while the NLOS results are coloured red. Most of the measurements with a distance shorter than 25\,m between Tx and Rx had a clear LOS with only a few measurements in NLOS condition between 12 and 25\,m separation. While the antenna position dictates if the channel is LOS or NLOS, analysis of the LOS and NLOS path losses separately have shown that the antenna height does not have a significant impact on the path loss.

Two path loss models, namely the floating intercept (FI) and the fixed reference (FR) model according to (\ref{eq:pathloss}) and (\ref{eq:pathloss_fr}) are considered:
\begin{equation}
    PL_{\mathrm{FI}}(d) = PL_0(d_0) + 10 n \log_{10} \left(\frac{d}{d_0} \right) + X_\sigma
    \label{eq:pathloss}
\end{equation}
\begin{equation}
    PL_{\mathrm{FR}}(d) = PL_{\mathrm{fs}}(d_0) + 10 \bar{n} \log_{10} \left(\frac{d}{d_0} \right) + X_{\bar{\sigma}},
    \label{eq:pathloss_fr}
\end{equation}
where $d_0$ is the reference distance, $d$ is the Tx-Rx distance (3D), $PL_0(d_0)$ and $PL_\mathrm{fs}(d_0)$ are the modelled FI path loss and, respectively, the FSPL at reference distance $d_0$, $n$ and $\bar{n}$ are the path loss exponents (PLE), and $X_\sigma$ and $X_{\bar{\sigma}}$ are the lognormal random shadowing variables with 0\,dB mean and standard deviation $\sigma$ and $\bar{\sigma}$. 
The parameters derived from least squares fitting are shown in Table \ref{tab:pathloss} and visualized in Fig. \ref{fig:pathloss}. The LOS models have an almost constant offset to the FSPL of about 3\,dB which can be explained by additional received power due to multipath components. The FI model for NLOS shows that the impact of an obstructed LOS path is higher at shorter distances. At larger distances, the NLOS model values come closer to the LOS values. Compared to the ray tracing results of 
\cite{Solomitckii2018} with $PL_0(1\,\textrm{m})=54.9\,\textrm{dB}$, $\mathit{n}=2.1$ for LOS, the results for the FI model in this paper show a 4.8\,dB higher $PL_0(1\,\textrm{m})$ with a similar $\mathit{n}$. The NLOS model in \cite{Solomitckii2018} with $PL_0(1\,\textrm{m})=24.6\,\textrm{dB}$ and $\mathit{n}=5.3$ is not a good fit to the results in this paper.

\vspace{-1em}
\begin{table}[ht]
\renewcommand{\arraystretch}{1.0}
\caption{Path loss model parameters for $d_0=1\,\mathrm{m}$}
\vspace{-1em}
\label{tab:pathloss}
\centering
\begin{tabular}{|c|c|c|c|c|c|c|}
\hline
 & \multicolumn{3}{|c|}{FI model} &  \multicolumn{3}{|c|}{FR model} \\ 
 \hline
 & $PL_0(d_0)$  & $\mathit{n}$ & $\sigma$ & $PL_{\mathrm{fs}}(d_0)$ & $\bar{n}$ & $\bar{\sigma}$\\
 & (dB)         &              &  (dB)    &  (dB)                   &           & (dB) \\
\hline
LOS  & 59.7 & 1.9 & 1.2 & 61.4 & 1.8 & 1.2 \\
\hline
NLOS & 77.4 & 0.9 & 1.6 & 61.4 & 2.0 & 2.0 \\
\hline
\end{tabular}
\end{table}

\subsection{Blockage Loss}
\begin{figure}[htb]
    \centering
    \includegraphics[width=0.45\textwidth]{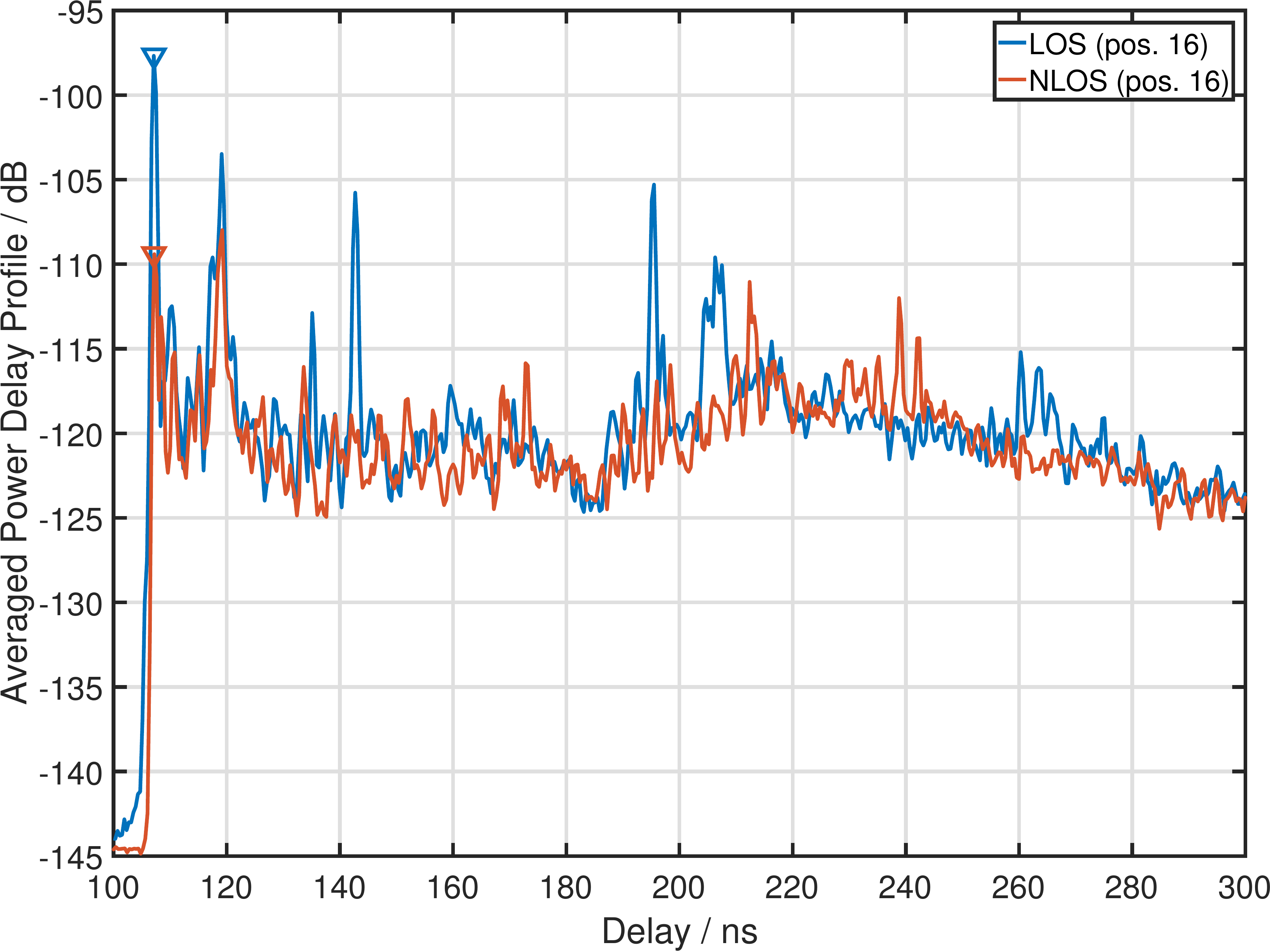}
    \vspace{-0.5em}
    \caption{Blockage Loss: Sample APDPs in LOS and NLOS condition}
    \label{fig:blockageloss}
\end{figure}

In an industrial environment, blockage losses (BL) are at least as important as path losses due to distance between Tx and Rx alone. These losses occur when, at a given position, the clear LOS gets interrupted by one or several machines, resulting in NLOS conditions. This happens, for example, when the antenna height changes. In order to evaluate blockage losses, the peak MPCs for measurement points where the high antenna position led to LOS condition and the low position to NLOS were compared. Figure \ref{fig:blockageloss} shows the APDPs for a single measurement point with both antenna heights. For LOS, the path gain of the main component (peak) is $-98$\,dB. In NLOS, the same component can still be identified but with $11$\,dB less path gain at $-109$\,dB. Two very prominent MPC peaks in LOS at 142 and 195 ns delay completely disappear into the DMCs in NLOS. The overall difference in path loss due to blockage at this measurement point adds up to 3.5\,dB.

\begin{figure}[htb]
    \centering
    \includegraphics[width=0.45\textwidth]{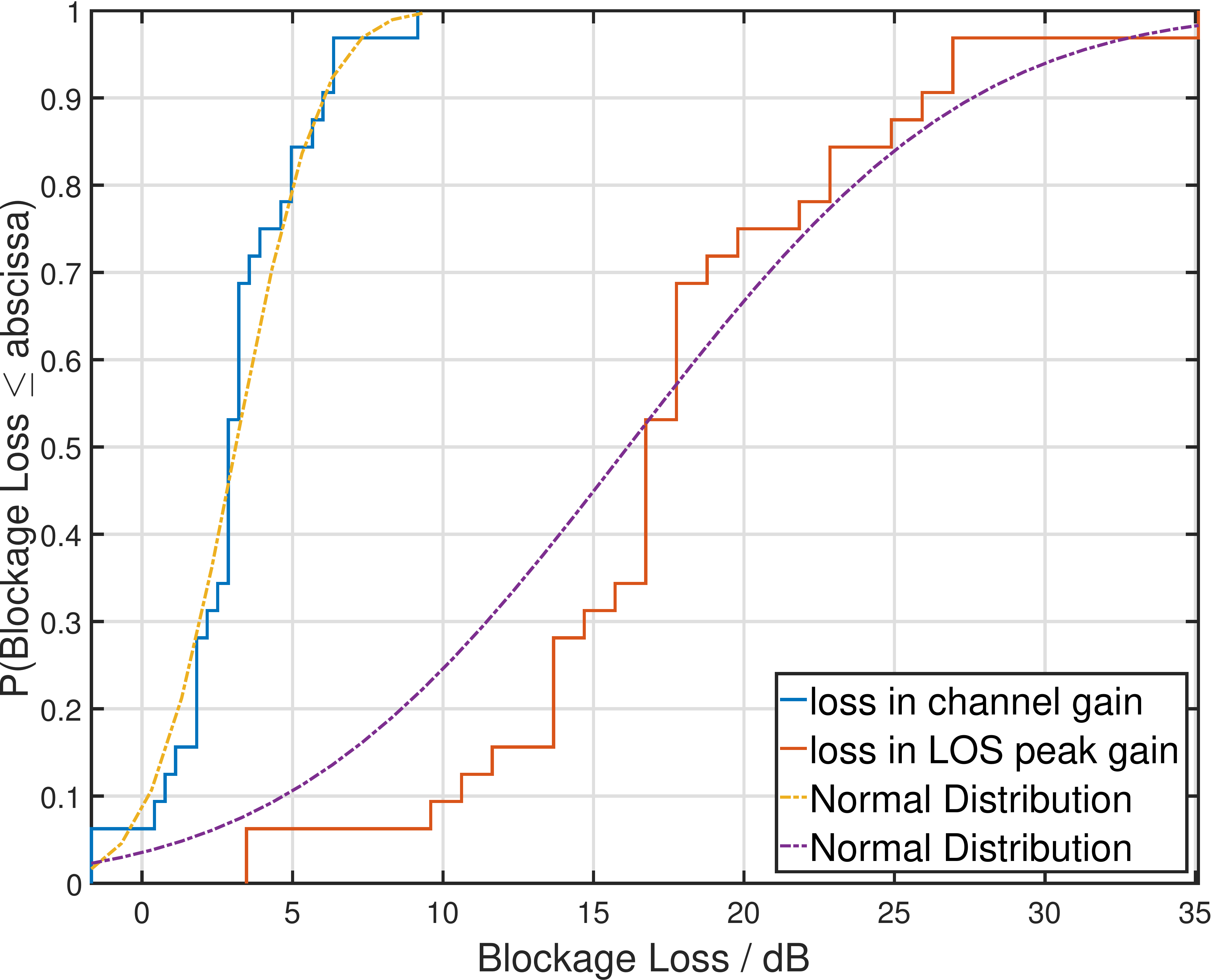}
    \vspace{-0.5em}
    \caption{CDF of blockage loss due to LOS/NLOS transition; Normal distributions parameterized accordingly to fit the blockage loss CDFs}
    \label{fig:bl_cdf}
\end{figure}

The cumulative distribution functions (CDF) of the differences in overall channel gain and LOS peak power due to blockage are shown in Fig. \ref{fig:bl_cdf}, together with fitted models.

The mean difference in overall channel gain is $3.11$\,dB with a standard deviation of $2.25$\,dB while the mean difference in LOS peak power is $16.14$\,dB with a standard deviation of $8.93$\,dB. This shows that, even though the LOS component is highly attenuated when blocked, the overall impact of blockage on the channel gain is limited.

\subsection{RMS Delay Spread}
\begin{figure}[htb]
    \centering
    \includegraphics[width=0.45\textwidth]{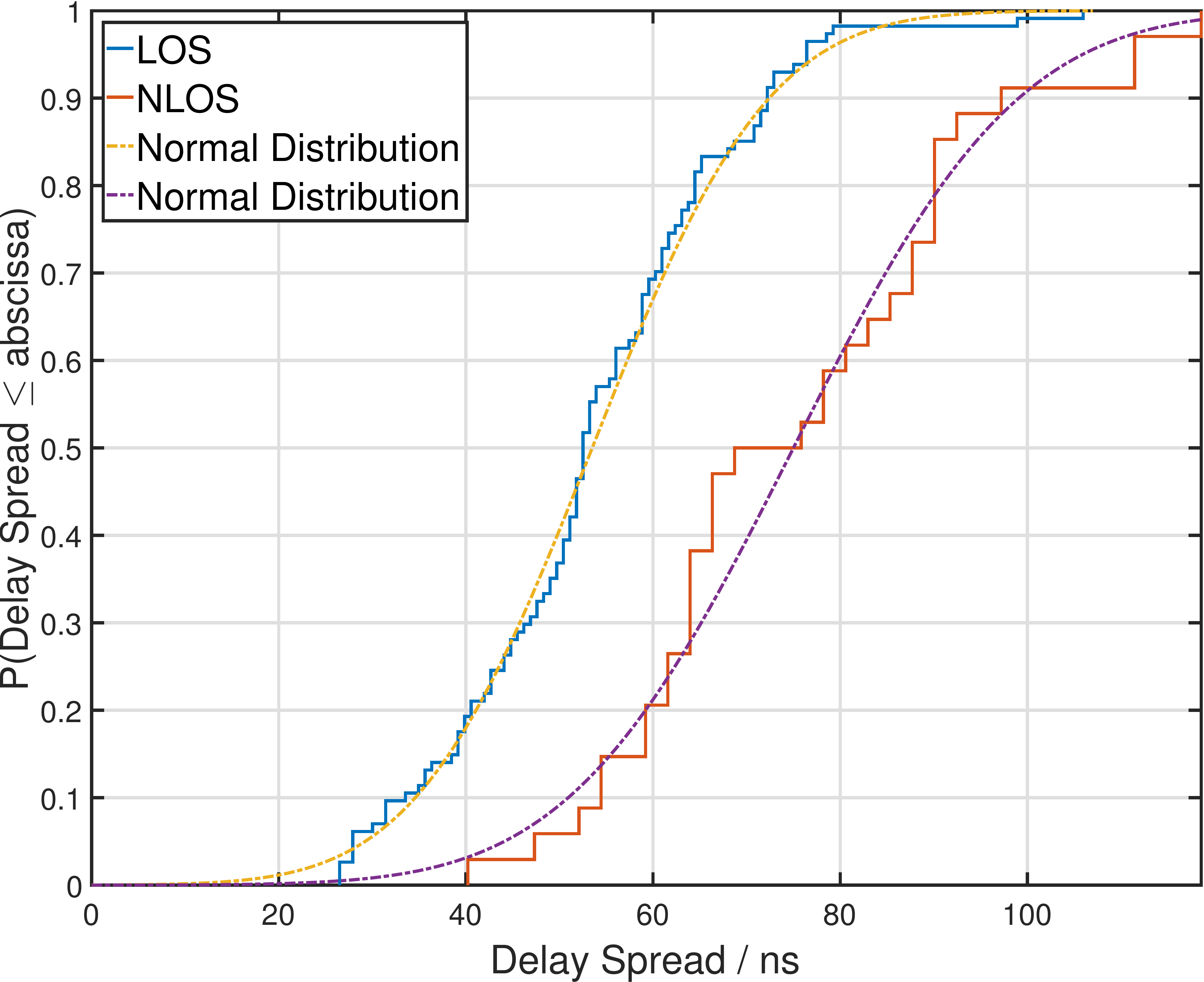}
    \vspace{-0.5em}
    \caption{CDF of RMS delay spread in LOS and NLOS condition; Normal distributions parameterized accordingly to fit the DS CDFs}
    \label{fig:ds_cdf}
\end{figure}

Based on the APDPs, the root mean square (RMS) delay spread (DS) was calculated. A relative evaluation threshold of 40\,dB was applied. CDFs of the DS for LOS and NLOS are illustrated in Fig. \ref{fig:ds_cdf}. They show a relatively constant offset of approximately $20$ ns with the spread for NLOS being higher than for LOS. Table \ref{tab:delayspread} shows the statistical parameters: mean $\mu_{DS}$, median $m_{DS}$, standard deviation $\sigma_{DS}$ and 95\%-quantile $Q_{DS.95}$. In order to be able to compare the statistical parameters with the 3GPP channel model, the values are also given on a logarithmic scale as defined in \cite{3GPP-TR-38901}.

\begin{table}[ht]
\vspace{-2em}
\renewcommand{\arraystretch}{1.0}
\caption{Statistical parameters of the RMS DS with 3GPP Indoor Office \cite{3GPP-TR-38901} for comparison}
\vspace{-1em}
\label{tab:delayspread}
\centering
\begin{tabular}{|c|c|c|c|c|}
\hline
 & $\mu_{DS}$ (ns) & $m_{DS}$ (ns) & $\sigma_{DS}$ (ns) & $Q_{DS.95}$ (ns) \\
\hline
LOS  & 53.6 & 52.5 & 14.8 & 76.5\\
\hline
NLOS & 75.3 & 71.4 & 18.8 & 111.8 \\
\hline\hline
 & $\mu_{lgDS}$ & 3GPP $\mu_{lgDS}$ & $\sigma_{lgDS}$ & 3GPP $\sigma_{lgDS}$ \\
\hline
LOS  & -7.29 & -7.71 & 0.12 & 0.18 \\
\hline
NLOS & -7.14 & -7.58 & 0.11 & 0.20 \\
\hline
\end{tabular}
\end{table}

Compared to the findings of Rappaport \cite{Rappaport89a}, where the median RMS delay spread was found to be 96\,ns for LOS and 105\,ns for obstructed LOS at 1.3\,GHz, the results of this measurement campaign show a much lower RMS delay spread. This can be explained by the higher attenuation of multipath components at 28\,GHz compared to 1.3\,GHz, and is supported by the ray tracing results of Solomitckii et al. \cite{Solomitckii2018} with a mean RMS delay spread of 38.5\,ns for LOS and 49.4\,ns for NLOS in the H28 scenario.

Table \ref{tab:delayspread} lists the corresponding 3GPP TR 38.901 \cite{3GPP-TR-38901} channel model parameters. Comparison of both the mean values ($\mu_{lgDS}$) and standard deviations ($\sigma_{lgDS}$) show that the \textit{Indoor-Office} scenario is not a good fit for the considered industrial scenario. The evaluated mean values are 0.42 (LOS) and 0.44 (NLOS) higher and the standard deviations are 0.06 (LOS) and 0.09 (NLOS) lower than the 3GPP model. Further comparison with the 3GPP model show that the LOS mean value lies between the \textit{Urban Micro-Street Canyon} (UMi-Street Canyon) scenario with $-7.49$ and the \textit{Urban Macro} (UMa) scenario with $-7.09$. The NLOS mean value is similar to the UMa scenario with $-7.18$.

\subsection{Angular Spread}
\begin{figure}[ht]
    \centering
    \includegraphics[width=0.35\textwidth]{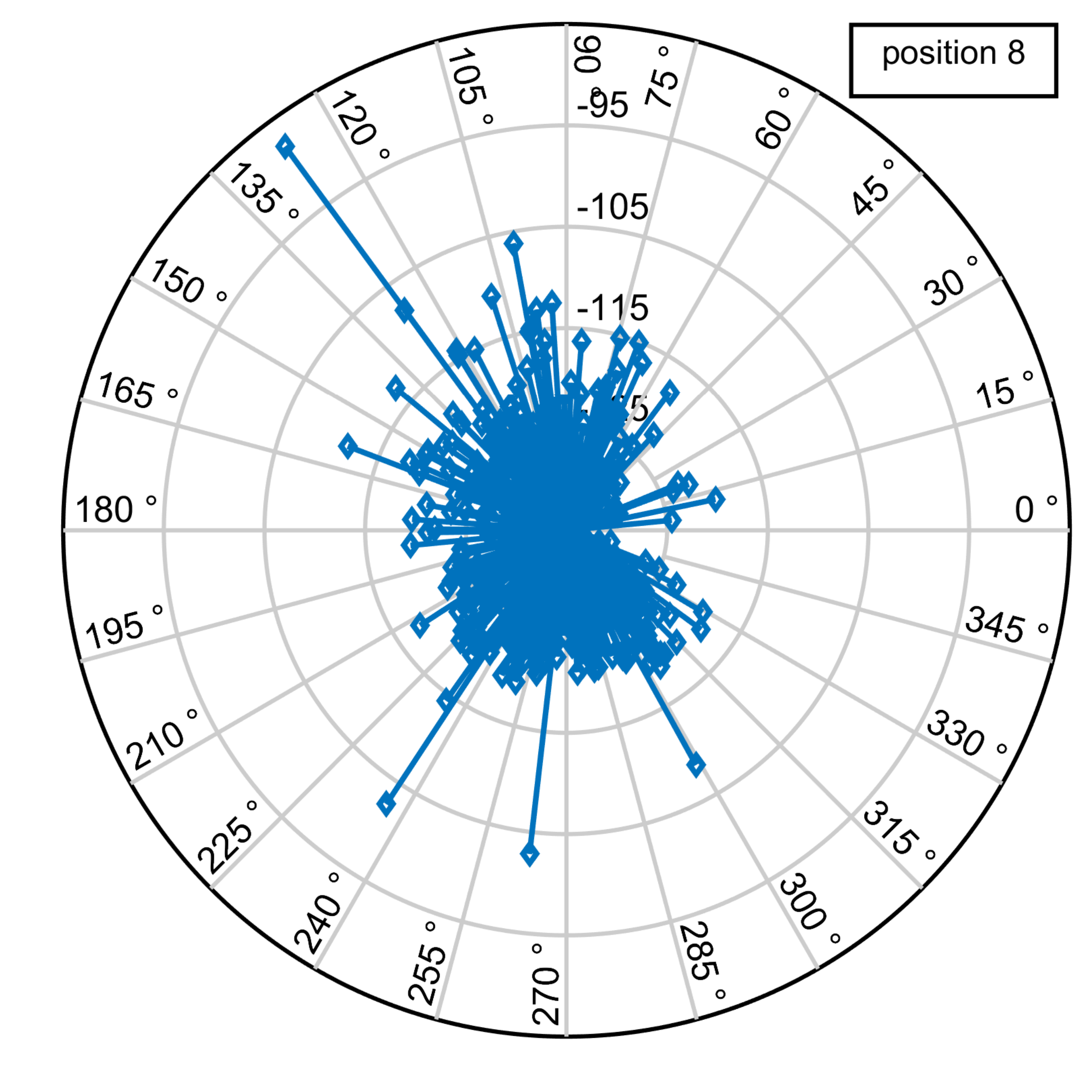}
    \vspace{-1.1em}
    \caption{Angular power profile in azimuth for one measurement point}
    \vspace{-1.1em}
    \label{fig:app}
\end{figure}

Direction-of-arrival (DoA) and direction-of-departure (DoD) information were extracted for each of the 162 measurements by applying the RB-MUSIC algorithm to the CIR snapshots. Based on the DoA and DoD information, angular power profiles (APPs) were estimated and the azimuth angular spread of arrival (ASA) and departure (ASD) were evaluated. Fig. \ref{fig:app} shows the APP in azimuth for a single measurement point for LOS. The LOS path can clearly be identified coming in from $128^{\circ}$  with strong reflections from $300$, $260$ and $235^{\circ}$. Dense multipath components can be seen impinging from all directions with a small gap between $345$ and $0^{\circ}$.
\vspace{-1.1em}

\begin{figure}[htb]
    \centering
    \includegraphics[width=0.45\textwidth]{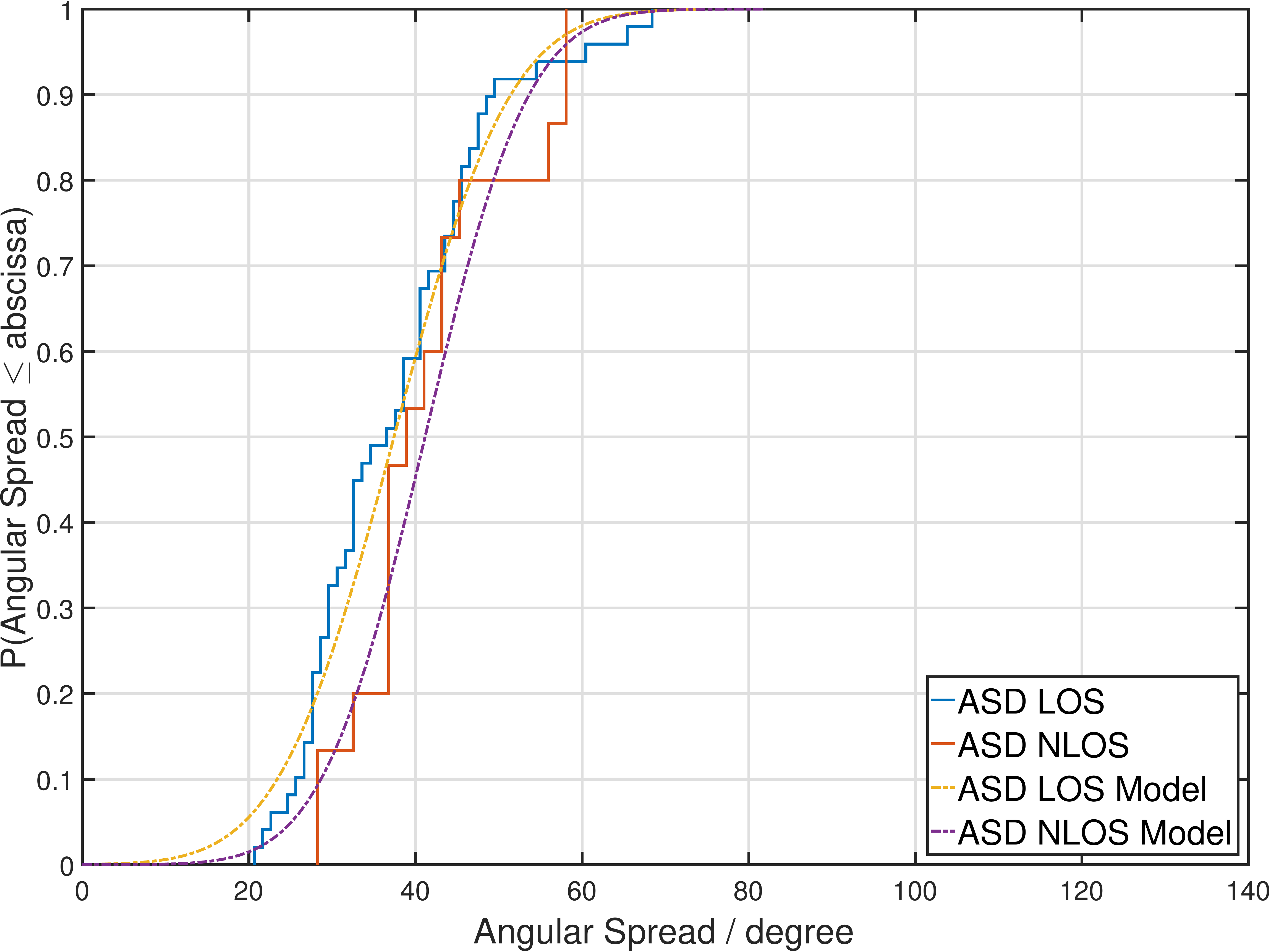}
    \vspace{-0.5em}
    \caption{CDF of RMS ASD in LOS and NLOS condition; Normal distributions parameterized accordingly to fit the ASD CDFs}
    \label{fig:app_cdf_pos0}
\end{figure}

\begin{figure}[htb]
    \centering
    \includegraphics[width=0.45\textwidth]{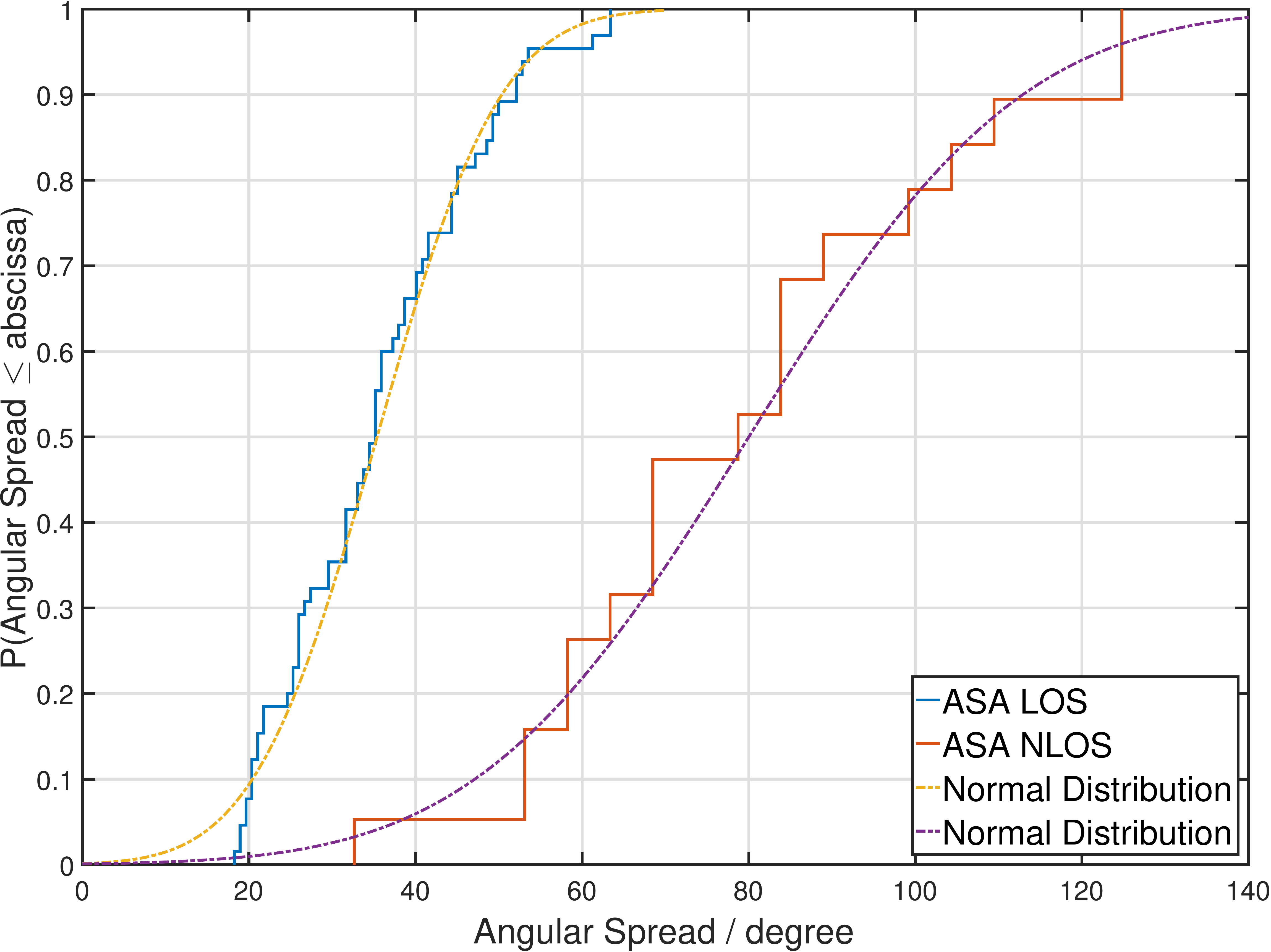}
    \vspace{-0.5em}
    \caption{CDF of RMS ASA in LOS and NLOS condition; Normal distributions parameterized accordingly to fit the ASA CDFs}
    \vspace{-1em}
    \label{fig:app_cdf_pos1}
\end{figure}

In Fig. \ref{fig:app_cdf_pos0} and \ref{fig:app_cdf_pos1}, the RMS azimuth angular spread of arrival and departure, respectively, are illustrated for both LOS and NLOS. The angles of arrival were evaluated for the UE positions, the angles of departure from the BS position. As with the delay spread, the ASA is higher in NLOS than in LOS. In contrast, the ASD is both lower and similar in LOS and NLOS. An explanation for this effect is that the scattering objects that lead to a high delay spread are localized around the UE positions. The statistical parameters are given in Table \ref{tab:angularspread} together with 3GPP \textit{Indoor-Office} values for comparison. Both ASA and ASD are of the same order of magnitude as the 3GPP \textit{Indoor-Office} values. While the ray tracing results in \cite{Solomitckii2018} with an ASA of 58.7\,{\degree} in LOS and 53.3\,{\degree} for NLOS  and an ASD of 43.1\,{\degree} for LOS and 65.3\,{\degree} for NLOS in the H28 scenario are not a perfect fit to the results of this measurement campaign, they are in a similar range. In contrast to the results in this paper, the LOS/NLOS condition impacts the ASD much more than the ASA.

\begin{table}[hbt]
\vspace{-1em}
\renewcommand{\arraystretch}{1.0}
\caption{Statistical parameters of ASA and ASD with 3GPP Indoor Office \cite{3GPP-TR-38901} for comparison}
\vspace{-1em}
\label{tab:angularspread}
\centering
\begin{tabular}{|c|c|c|c|c|}
\hline
%
%
 & $\mu_{ASA}$ (\degree) & $m_{ASA}$ (\degree) & $\sigma_{ASA}$ (\degree)& $Q_{ASA.95} (\degree)$ \\
\hline
LOS  & 35.38 & 34.96 & 11.67 & 55.25\\
\hline
NLOS & 80.01 & 81.09 & 25.66 & 126.64 \\
\hline%
%
%
\hline%
 & $\mu_{lgASA}$ & 3GPP $\mu_{lgASA}$ & $\sigma_{lgASA}$ & 3GPP $\sigma_{lgASA}$ \\
\hline
LOS  & 1.53 & 1.50 & 0.15 & 0.29 \\
\hline
NLOS & 1.88 & 1.70 & 0.15 & 0.23 \\
\hline%
%
%
\hline%
 & $\mu_{ASD}$ (\degree) & $m_{ASD}$ (\degree) & $\sigma_{ASD}$ (\degree)& $Q_{ASD.95}$ (\degree)\\
\hline
LOS  & 37.41 & 36.37 & 10.93 & 60.27\\
\hline
NLOS & 41.14 & 38.73 & 9.74 & 58.63 \\
\hline%
%
%
\hline%
& $\mu_{lgASD}$ & 3GPP $\mu_{lgASD}$ & $\sigma_{lgASD}$ & 3GPP $\sigma_{lgASD}$ \\
\hline%
LOS  & 1.56 & 1.60 & 0.12 & 0.18 \\
\hline
NLOS & 1.60 & 1.62 & 0.10 & 0.25 \\
\hline
\end{tabular}
\end{table}


\section{Conclusion}
In this paper, we presented the channel sounder setup, scenario and first results of a wideband channel measurement campaign in a machine hall at 28 GHz. Evaluation of the power delay profiles shows that the radio channel is not sparse, but contains dense multipath components (DMC) with a delay of up to 1000 ns. The ratio between MPC and DMC still has to be evaluated. Based on the measurement results, a path loss model in LOS and NLOS condition was parameterized and blockage losses due to transition from LOS to NLOS were assessed. While blockage has a high impact on the LOS component, the overall influence on channel gain is limited. Evaluation of the delay and angular spread and comparison to the 3GPP TR 38.901 channel model show that none of the scenarios specified by 3GPP are a good fit for industrial environments, emphasizing the need for a specific scenario and parameter set. As the 3GPP GSCM does not support DMC with additional specular components, an extension to the channel model is necessary.  Compared to the ray tracing results of Solomitckii et al. \cite{Solomitckii2018} in the 28\,GHz \textit{Heavy Industry} scenario, the results presented in this paper are of the same order of magnitude but still differ considerably. This highlights the need for proper scenario definition and calibration of ray tracers with measured data.

\vspace{5pt}
\section*{Acknowledgment}
The authors would like to thank Fraunhofer IPK for the great support and the opportunity to carry out the measurement campaign in the machine hall of the Production Technology Center (PTZ) Berlin.

This work has been supported by the Governing Mayor of Berlin, Senate Chancellery -- Higher Education and Research, and the European Union -- European Regional Development Fund.



\bibliographystyle{IEEEtran}

%
%



\end{document}